\documentclass[published]{JHEP3} 

\JHEP{00(2009)000}

\JHEPspecialurl{http://jhep.sissa.it/JOURNAL/JHEP3.tar.gz}

\usepackage{epsfig,multicol,bbm}

\newcommand\fverb{\setbox\fverbbox=\hbox\bgroup\verb}
\newcommand\fverbdo{\egroup\medskip\noindent%
            \fbox{\unhbox\fverbbox}\ }
\newcommand\fverbit{\egroup\item[\fbox{\unhbox\fverbbox}]}
\newbox\fverbbox

\title{Signature of short distance physics on inflation power spectrum and CMB anisotropy }%

\author{Suratna Das \\
    Physical Research Laboratory, Ahmedabad, India, 380009\\
    E-mail: \email{suratna@prl.res.in}\\ }
\author{Subhendra Mohanty\\
    Physical Research Laboratory, Ahmedabad, India, 380009\\
        E-mail: \email{mohanty@prl.res.in}}%

\received{\today}
\accepted{}     

\abstract{ The inflaton field responsible for inflation may not be a
  canonical fundamental scalar. It is possible that the inflaton is a
  composite of fermions or it may have a decay width. In these cases
  the standard procedure for calculating the power spectrum is not
  applicable and a new formalism needs to be developed to determine
  the effect of short range interactions of the inflaton on the power
  spectrum and the CMB anisotropy. We develop a general formalism for
  computing the power spectrum of curvature perturbations for such
  non-canonical cases by using the flat space K\"{a}ll\'{e}n-Lehmann
  spectral function in curved quasi-de Sitter space assuming
  implicitly that the Bunch-Davis boundary conditions enforces the
  inflaton mode functions to be plane wave in the short wavelength
  limit and a complete set of mode functions exists in quasi-de Sitter
  space. It is observed that the inflaton with a decay width
  suppresses the power at large scale while a composite inflaton's
  power spectrum oscillates at large scales. These observations may be
  vindicated in the WMAP data and confirmed by future observations
  with PLANCK.}


\begin{document}

\section{Introduction}

In the generic inflation model \cite{guth}, inflation is caused by a
slow roll of the inflaton scalar field and the perturbations of the
inflaton field give rise to density perturbations \cite{guth2} and CMB
anisotropies observed at cosmological scales.  The two-point
correlation function of the inflaton perturbation during inflation or
the power spectrum of this two-point correlation in momentum space
determines the CMB anisotropy of the universe at last scattering which
we observe today. The generic slow roll model of inflation are
characterized by the inflaton potential and its derivatives and a
large variety of particle physics potential have been studied
\cite{lyth-riotto}. Inflation may be caused by more than one scalar
field and these multifield models have interesting consequences in the
CMB anisotropy like isocurvature perturbation \cite{gordon} or large
non-gaussianity in curvaton models \cite{curvaton}. In models of
inflation with elementary scalar fields, the perturbations of the
inflaton obey the Klein-Gordon equation in the quasi de Sitter space
\cite{riotto}, whose solutions are used for calculation of the two
point correlation and the curvature power spectrum.

However, it may be possible that the inflaton field is a composite of
fermions and we can ask if the compositeness changes the perturbation
spectrum which can be observed in the CMB anisotropy.  Similarly if
the inflaton is unstable with a decay width $\Gamma \sim H/N$ (such
that the inflaton decays after N e-foldings of inflation are over)
then again we can ask if the decay of the inflaton is reflected in the
power spectrum and CMB anisotropy. For such situations the standard
methods of calculating the power spectrum do not work as not all forms
of the short range structure of the scalar field are reflected in the
inflaton potential. If the length scale of the scalar perturbation is
of the same order as the compositeness scale then the effective theory
description of the scalar potential breaks down. Similarly if the
inflaton is a resonance with a lifetime of the same order as the
duration of inflation $\tau \sim N/H$ i.e a width $\Gamma \sim H/N$
then there are corrections to the two point correlation that are not
reflected in the inflaton potential.

In this paper we find a general method for computing the power
spectrum of inflaton perturbations if the inflaton has non-trivial
interactions like a decay width or if the inflaton is not an
elementary scalar but a composite of fermions. \emph{We show in
  general that the two-point correlations of the interacting field can
  be written in terms of the two-point function of the free field (in
  the de Sitter background) by use of the K\"{a}ll\'{e}n-Lehmann
  spectral function \cite{Kallen,lehman} if the short wavelength limit
  of the mode functions are plane wave states
  $\frac{1}{\sqrt{2k}}e^{-ik\eta}$, which in the quasi de-Sitter space
  is enforced by the assumption of the Bunch-Davis boundary conditions
  and there exists a complete orthonormal set of mode functions of the
  free theory in curved spacetime which is true in the quasi de-Sitter
  space relevant for inflation power spectrum calculation.} The two
point correlation of an interacting theory can be written as a
convolution of the free field correlation function with a spectral
function $\rho(\sigma^2)$
\begin{equation}
G^{\left({\rm int}\right)}(p)=\int_0^{\infty} d\sigma^2 \rho(\sigma^2)
G^0(p,\sigma^2),
\end{equation}
where $G^{\left({\rm int}\right)}(p)$ is the two point correlation of
the interacting theory and $G^0(p,\sigma^2)$ is the free-field
correlation with mass parameter $\sigma$. The K\"{a}ll\'{e}n-Lehmann
representation holds for all two point correlations like the Feynman
propagator $\Delta(p,\sigma^2)$ or the equal time Wightman function
$W_{\rm ET}(x-y)$. As mentioned above we show that this result can be
generalized to the curved space if we assume that a complete
orthogonal basis set of states of the interacting theory exists in
curved spacetime.

The power spectrum of the inflaton perturbation is related to the
equal-time Wightman function in the de-Sitter space as
\begin{eqnarray}
W^{\rm dS}_{\rm ET}(x)= \langle0|(\delta\phi({\mathbf
x},t))^2|0\rangle=\int\frac{dk}{k}{\mathcal P}_{\delta\phi}(k).
\label{var1}
\end{eqnarray}
The Bunch-Davies boundary condition is that the inflaton
perturbations, in the limit where the momentum $k$ is large compared
to the inflaton horizon (for a spatially flat de-Sitter space), tend
to the free field form $\delta \phi(k,\eta)=\frac{1}{\sqrt{2 k}} e^{-i
  k \eta}$ (where $\eta$ is the conformal time). Assuming the
Bunch-Davies boundary conditions, if we have short range interactions
which dominate at scales smaller than the inflation horizon, we may be
justified in using the flat space form of the spectral function in
Eq.~(\ref{var1}) to compute the two point correlation function for
interacting theory. Therefore power spectrum of the interacting scalar
field can be expressed as
\begin{equation}
P^{\rm
\left(int\right)}(k)=\int_0^{\infty}P^{\left(0\right)}(k,\sigma^2)\rho\left(\sigma^2\right)d\sigma^2,
\label{power1}
\end{equation}
where $P^{\left(0\right)}(k,\sigma^2)$ is the power spectrum of the
free scalar field with a mass parameter $\sigma$ and
$\rho(\sigma^2)$ is the KL spectral function which encapsulates all
the short distance interactions (like compositeness or resonance) of
the scalar field.

In Sec.~(\ref{power-interacting}) we derive the relation between the
power spectrum of interacting theory and the free field theory given
in Eq.~(\ref{power1}). In Sec.~(\ref{case:decay}) we apply this result
to calculate the power spectrum for case of a decaying inflaton
field. We find that the TT angular spectrum of CMB anisotropy is
suppressed at low $l$. In Sec.~(\ref{case:composite}) we derive the
power spectrum of the composite inflaton field. We find that the power
spectrum of the composite field has a resonance which gives rise to
oscillatory features in the TT angular spectrum.

\section{Power spectrum of interacting scalar field - general case}
\label{power-interacting}

The power spectrum for the inflaton is essentially given by the
equal-time Wightman function in de-Sitter space. In this section we
will provide a general formalism of calculating power spectrum for
interacting scalar field using KL representation. Derivations of the
two point correlation functions for interacting real scalar field
using KL representation in Minkowski space is given in
Appendix~(\ref{KL}).

It is assumed in the following derivation that the asymptotic `in' and
`out' states of an interacting scalar field are free particle states
in the curved space. We assume the interactions being short ranged
dominate over curvature effects at short distances. Since we assume
the Bunch-Davies boundary conditions that the curved space mode
functions in the large momentum limit go over to the flat space
plane-wave form, we can directly use the flat-space calculation of
spectral function of the interaction theory in the inflation power
spectrum formula.

To generalize the KL formalism in de-Sitter space it is to be noted
that in de-Sitter space there is no translational invariance in the
time direction like Minkowski space. So, the mode functions given in
Eq.~(\ref{mode}) can be written in a more general form for the
inflaton fluctuations as
\begin{eqnarray}
\langle0|\delta\phi(x)|n\rangle&=&\left(\sqrt{2p^0_n}\right)\delta\phi(p^0_n,\eta){\rm e}^{i\mathbf{p}_n\cdot \mathbf{x}}\langle0|\delta\phi(0)|n\rangle,
\end{eqnarray}
where $\delta\phi(p^0_n,\eta)$ are the free field mode functions
which obey the Klein-Gordon equation in the curved background and in
the flat space limit $\delta\phi(p^0_n,\eta)=
\frac{1}{\sqrt{2p^0_n}}\exp{\left(-ip^0_n\eta\right)}$. In
Appendix~(\ref{power}) we derive the explicit form for the mode
functions of a massive scalar in the de-Sitter space.

Hence the Wightman function in de-Sitter space can be written as
\begin{eqnarray}
W^{\rm dS}_{\rm ET}(x,y)=\langle0|\delta\phi(x)\delta\phi(y)|0\rangle=\sum_n\left(2p^0_n\right)\delta\phi(p^0_n,\eta)\delta\phi(p^0_n,\eta^{\prime}){\rm e}^{i\mathbf{p}_n\cdot\left(\mathbf{x}-\mathbf{y}\right)}|\langle0|\delta\phi(0)|n\rangle|^2.
\label{phi_phi}
\end{eqnarray}
Here $\langle0|\delta\phi(0)|n\rangle$ represents the short range
interactions of the interacting inflaton perturbations and according
to our previous assumption can be replaced by the spectral function
$\rho(q^2)$ of Minkowski space defined in Eq.~(\ref{sf}) as
\begin{eqnarray}
\theta(q^0)\rho(q^2)=(2\pi)^3\sum_n\delta^4(q-p_n)|\langle0|\Phi(0)|n\rangle|^2.
\label{sf_1}
\end{eqnarray}
With this definition of spectral function Eq.~(\ref{phi_phi}) can be written as
\begin{eqnarray}
\langle0|\delta\phi(x)\delta\phi(y)|0\rangle
&=&\int\frac{d^4q}{\left(2\pi\right)^3}\int_0^{\infty}d\sigma^2\left(2q^0\right)\delta\phi(q^0,\eta)\delta\phi(q^0,\eta^{\prime}){\rm e}^{i\mathbf{q}\cdot\left(\mathbf{x}-\mathbf{y}\right)}\theta(q^0)\rho(\sigma^2)\delta(q^2+\sigma^2)\nonumber \\
&=&\int_0^{\infty}d\sigma^2\rho(\sigma^2)\int\frac{d^3q}{\left(2\pi\right)^3}\delta\phi(\omega,\eta)\delta\phi(\omega,\eta^{\prime}){\rm e}^{i\mathbf{q}\cdot\left(\mathbf{x}-\mathbf{y}\right)}.
\label{phi_phi-1}
\end{eqnarray}
The equivalent form of Wightman function in Minkowski space of the
above equation is given in Eq.~(\ref{wf}). Here
$\omega=\sqrt{\mathbf{q}^2+\sigma^2}$ and in de-Sitter space
$\delta\phi(\omega,\eta)$ has the solution given in Eq~(\ref{sol-phi})
with mass $m$ of the inflaton field is replaced by the mass parameter
$\sigma$. The solution for light scalar field given in
Eq.~(\ref{light}) will be used in the following derivation because
for very massive fields ($m_\phi>H$) the power spectrum is highly damped in
superhorizon scales as given in Eq.~(\ref{heavy}) and hence the upper
limit of $\sigma^2$ integration in the above equation should have a
cut-off at $m_0^2$ where $m_0\ll H$.

The equal-time Wightman function in de-Sitter space $W^{\rm dS}_{\rm
  ET}(x)$ gives the power spectrum for the inflaton fluctuations
\begin{eqnarray}
\langle0|\left(\delta\phi(x)\right)^2|0\rangle&=&\int_0^{m_0^2}d\sigma^2\rho(\sigma^2)\int\frac{dq}{q}\frac{q^3}{2\pi^2}\left|\delta\phi(\omega,\eta)\right|^2\nonumber \\
&=&\int\frac{dq}{q}\int_0^{m_0^2}d\sigma^2\rho(\sigma^2){\mathcal
P_{\delta\phi}^{\left(0\right)}}(q,\sigma^2), \label{dphi}
\end{eqnarray}
where ${\mathcal P_{\delta\phi}^{\left(0\right)}}(q,\sigma^2)$ is
the power spectrum of the free inflaton field given by
Eq.~(\ref{light}) with $m$ replaced by $\sigma$
\begin{eqnarray}
{\mathcal
P}^{\left(0\right)}_{\delta\phi}(q,\sigma^2)=\frac{H^2}{4\pi^2}\left(\frac{q}{2aH}\right)^{\frac23\frac{\sigma^2}{H^2}}.
\label{light_1}
\end{eqnarray}
Following Eq.~(\ref{var}) the power spectrum for the interacting
scalar field is given by
\begin{eqnarray}
\langle0|(\delta\phi({\mathbf
x},t))^2|0\rangle=\int\frac{dq}{q}{\mathcal P}^{\left({\rm
int}\right)}_{\delta\phi}(q). 
\label{var2}
\end{eqnarray}
From Eq.~(\ref{dphi}) and Eq.~(\ref{var2}) we get
\begin{eqnarray}
{\mathcal P}^{\left({\rm int}\right)}_{\delta\phi}(k)=\int_0^{m_0^2}d\sigma^2\rho(\sigma^2){\mathcal P_{\delta\phi}^{\left(0\right)}}(k,\sigma^2),
\end{eqnarray}
and hence the curvature power spectrum (defined in
Eq.~(\ref{curvature})) for interacting inflaton field will be
\begin{eqnarray}
{\cal P}_{\cal R}(k)=\frac{H^2}{\dot{\phi}^2}{\mathcal P}_{\delta\phi}^{\left({\rm int}\right)}(k) =\frac{1}{2m_{\rm Pl}^2\epsilon}\int_0^{m_0^2}{\mathcal P}_{\delta\phi}^{\left(0\right)}(k,\sigma^2)\rho\left(\sigma^2\right)d\sigma^2,
\label{cps}
\end{eqnarray}
where $\epsilon$ is the slow roll parameter of the inflaton. This form
of curvature spectrum will be used as input in CAMB \cite{camb} or
CMBFAST \cite{cmbfast} to determine the CMB anisotropy spectrum from a
given model of inflaton interactions.

\section{Inflaton with a decay width}
\label{case:decay}

From the fact that the inflation must end in reheating we expect that
the inflaton has couplings to other particles and it can decay into
lighter particles. The inflaton decay width must be smaller than $H
/N$ (where $N \simeq 100$ is the number of e-foldings needed to solve
the horizon and flatness problems). Since $\Gamma\lesssim 10^{-2} H$,
the decay width term is negligible compared to the $H\delta\dot{\phi}$
term in the Klein-Gordon equation given in Eq.~(\ref{KG}).

To compute the power spectrum of the decaying inflaton, we start
with the Breit-Wigner propagator in flat space, of an unstable
scalar particle with decay width $\Gamma$ and mass $m$
\begin{equation}
\Delta^{\left({\rm int}\right)}(q^2)=\frac{1}{q^2-m^2+im\Gamma},
\end{equation}
whose spectral function has the form \cite{decay},
\begin{eqnarray}
\rho(\sigma^2)=\frac{1}{\pi}\frac{m\Gamma}{\left(\sigma^2-m^2\right)^2+m^2\Gamma^2}.
\label{spec_d}
\end{eqnarray}
Using the spectral function from Eq~(\ref{spec_d}) in
Eq~(\ref{cps}) the power spectrum for inflaton with a decay width will
be
\begin{eqnarray}
{\cal P}_{\cal R}(k)&=&\frac{H^2}{8m_{\rm
    Pl}^2\epsilon\pi^2}\left[\tan^{-1}\left(\frac{m}{\Gamma}\right)-\tan^{-1}\left(\frac{m^2-m_0^2}{m\Gamma}\right)\right]+\frac{m^2}{12m_{\rm Pl}^2\epsilon\pi^2}\ln\left(\frac z2\right)\nonumber\\&\times&\left[\tan^{-1}\left(\frac
  m\Gamma\right)-\cot^{-1}\left(\frac{m\Gamma}{m^2-m_0^2}\right)+\frac{\Gamma}{2m}\ln\left(\frac{\left(m^2-m_0^2\right)^2+m^2\Gamma^2}{m^2\left(m^2+\Gamma^2\right)}\right)\right],
\end{eqnarray}
where $z=\frac{k}{aH}$ and $m_0\ll H$ is the cut-off scale for the
mass parameter $\sigma$.

In Fig~(\ref{fig_decay_pk}) we plot ${\cal P}_{\cal R}(k)$ vs. $k$
plot for the decaying inflaton. We observe that for the free scalar
field (i.e. $\Gamma=0$) the curvature power spectrum is
scale-invariant where for the decaying inflaton the power spectrum
gets suppressed at low $k$ and increases at high $k$ with respect to
the free inflaton case. We also observe that the higher the decay
width $\Gamma$, more is the suppression of power at low $k$ and
increase of power at high $k$. 
\FIGURE{ 
\centering
\includegraphics[angle=0,width=0.9\textwidth]{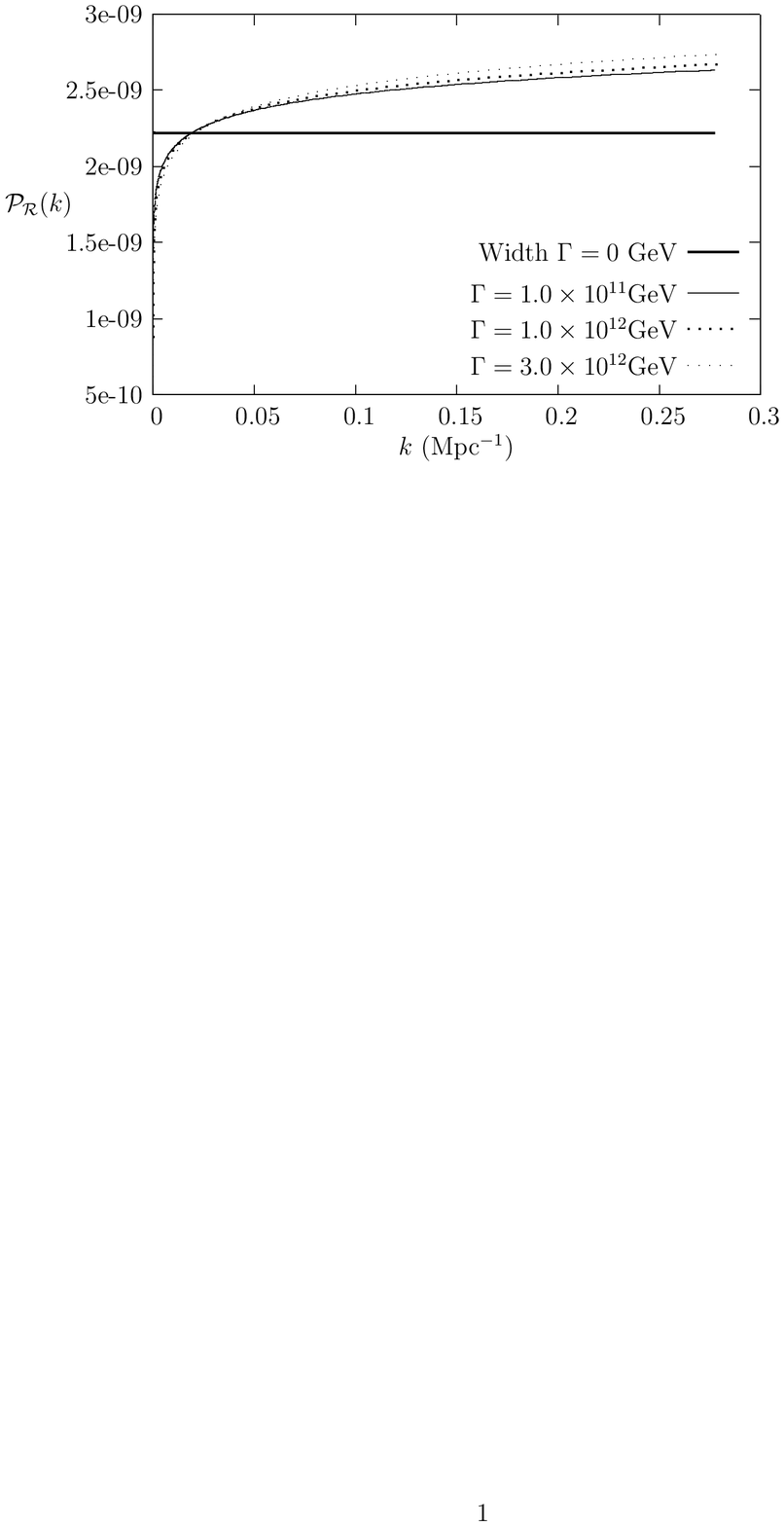}
\caption{${\cal P}_{\cal R}(k)$ vs. $k$ plot for decaying scalar
  inflaton}
\label{fig_decay_pk} }

In Fig~(\ref{fig_decay}) we plot the TT angular spectrum for the
inflaton with a decay width. The parameters used for the above plots
are $H=10^{13}$ GeV, $m=3.5\times 10^{12}$ GeV, $m_0=7.5\times
10^{12}$ GeV and for $\Gamma=1.0\times10^{11}$ GeV,
$\Gamma=1.0\times10^{12}$ GeV and $\Gamma=3.0\times10^{12}$ GeV the
values of $\epsilon$ used are $1.412\times 10^{-5},\, 1.29\times
10^{-5}$ and $1.069\times 10^{-5}$ respectively. In previous figure,
we find that as the inflaton decay width $\Gamma$ is increased the
power at large distance scales gets suppressed. This results in
suppression of the TT spectrum at low $l$ with increasing decay width
in this plot. A decay width of the inflaton may be a viable
explanation of the WMAP observation of suppression in the TT power
spectrum \cite{oscillations,decay-ref}.

\FIGURE{
\centering
\includegraphics[angle=270,width=1.0\textwidth]{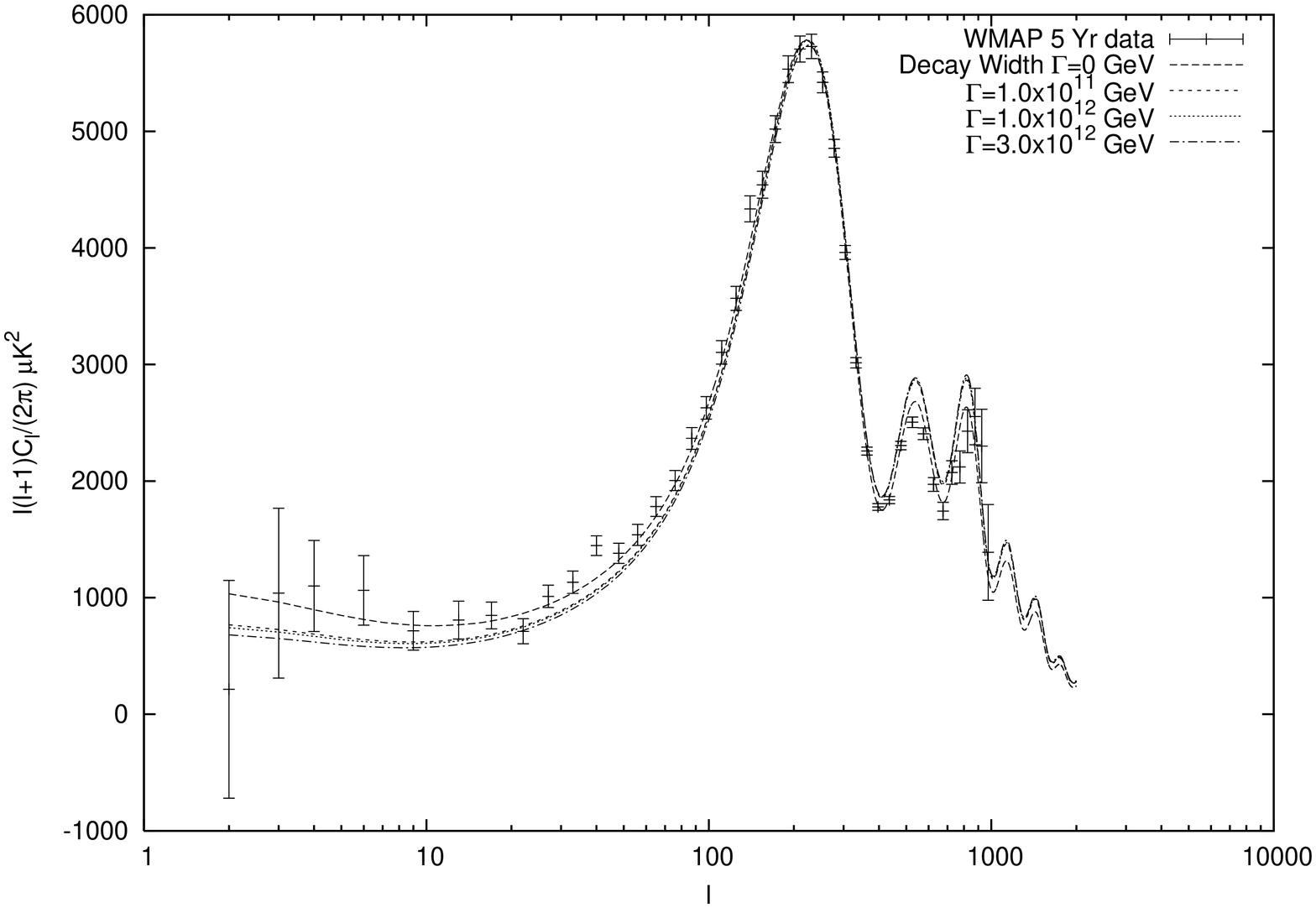}
\caption{The TT angular spectrum for the inflaton with a decay width.}
\label{fig_decay}
}

\section{Inflaton as Composite Particle}
\label{case:composite}

An interesting model of inflation can be with the inflaton as a GUT
scale techni-pion which arises from a condensate of fermions in a GUT
scale SU(N) techni-colour model \cite{tech} or the inflaton can be a
composite of heavy right handed neutrinos \cite{right}. In such models
one may ask in what way the compositeness of the inflaton affects the
power spectrum. We use the spectral representation of a composite
scalar in deriving the power spectrum.

The spectral function for a composite scalar can be
taken as in QCD models \cite{negele} as
\begin{equation}
\rho(\sigma^2)=Z\delta(\sigma^2-m_{\varphi}^2)+\frac{1}{f_{\varphi}^2m_{\varphi}^2}\rho_c(\sigma^2)\theta(\sigma^2-s_0^2),
\label{cssp}
\end{equation}
where $m_{\varphi}$ is the techni-pion mass, $f_{\varphi}$ is the
symmetry breaking scale and $s_0$ is the threshold for the onset of a
continuum contribution $\rho_c(\sigma^2)$.

The wave function renormalization constant $Z$ can be determined using
the following property of the spectral function
\begin{equation}
\int_0^{\infty}\rho(\sigma^2)d\sigma^2=1.
\label{z}
\end{equation}

The spectral function for the continuum is given as \cite{agnes}
\begin{equation}
\rho_c(\sigma^2)=\frac{N}{8\pi^2}\sigma^2\left(1-\frac{s_0^2}{\sigma^2}\right)^{\frac32},
\label{agnes_rho}
\end{equation}
where $N$ is the number of fermion flavours.
Using Eq.~(\ref{cssp}), Eq.~(\ref{z}) and Eq.~(\ref{agnes_rho}) we get
\begin{equation}
Z=1-\frac{N}{8\pi^2}\frac{1}{f_{\varphi}^2m_{\varphi}^2}\left(\frac 12\Lambda^4-\frac{3s_0^2}{2}\Lambda^2+s_0^4\right),
\label{Z_cal}
\end{equation}
where $\Lambda$ is the ultra-violet cut-off of the composite theory.

Now using Eq.~(\ref{cssp}) and Eq.~(\ref{agnes_rho}) in
Eq.~(\ref{cps}) we find the power spectrum for a composite scalar particle as
\begin{eqnarray}
{\cal P}_{\cal R}(k)&=&
\frac{ZH^2}{8\pi^2m_{\rm Pl}^2\epsilon}\left(\frac{z}{2}\right)^{\frac23\frac{m_{\varphi}^2}{H^2}}+\frac{3NH^4}{256\pi^4m_{\rm Pl}^2\epsilon\left[\ln\left(\frac z2\right)\right]^2}\frac{1}{f_{\varphi}^2m_{\varphi}^2}\left(\frac z2\right)^{\frac23\frac{s_0^2}{H^2}}\left[3H^2+s_0^2\ln\left(\frac z2\right)\right]\nonumber\\
&+&\frac{3NH^4}{256\pi^4m_{\rm Pl}^2\epsilon\left[\ln\left(\frac z2\right)\right]^2}\frac{1}{f_{\varphi}^2m_{\varphi}^2}\left(\frac z2\right)^{\frac23\frac{m_0^2}{H^2}}\left[-3H^2+\left(2m_0^2-3s_0^2\right)\ln\left(\frac z2\right)\right].
\end{eqnarray}

\FIGURE{
\centering
\includegraphics[angle=0,width=0.9\textwidth]{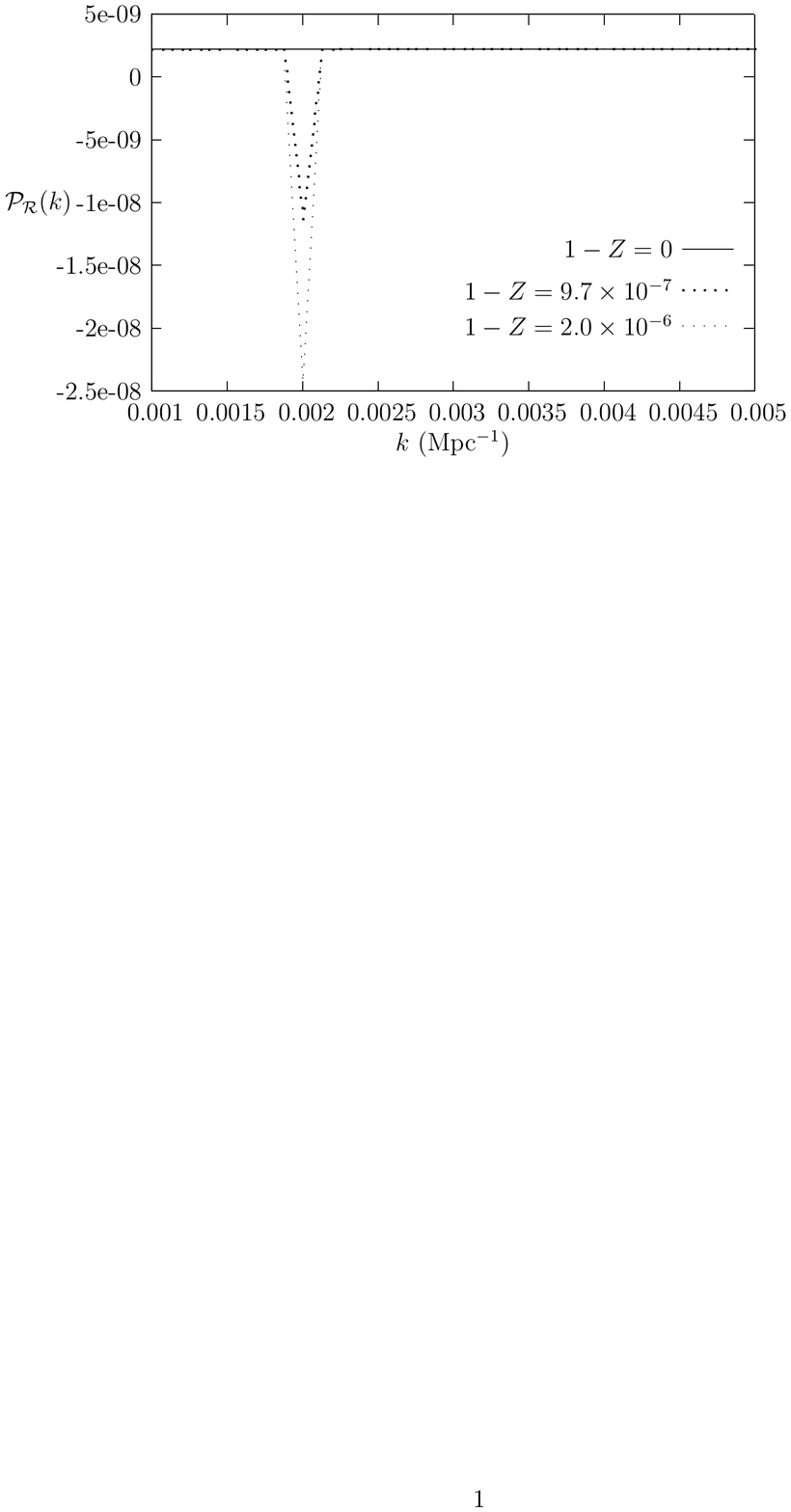}
\caption{${\cal P}_{\cal R}(k)$ vs $k$ plot for composite inflaton}
\label{fig_composite_pk}
}
In Fig.~(\ref{fig_composite_pk}) ${\cal P}_{\cal R}(k)$ vs $k$ plot
for composite inflaton is given. We see that though the curvature
power is scale invariant for a free scalar field (i.e. $1-Z=0$),
there is a sharp resonance at $k=0.002$ Mpc$^{-1}$, due to
compositeness $(1-Z > 0)$ in the inflaton field. The resonances
increases as the compositeness of the inflaton increases (smaller
$Z$). Such resonances in curvature power spectrum can lead to
oscillatory features in TT angular power of CMBR as seen in other
examples where spikes in the power spectrum can arise due a period
of fast roll \cite{Jain} or a bump in the potential \cite{Peiris}.


\FIGURE{
\centering
\includegraphics[angle=270,width=1.0\textwidth]{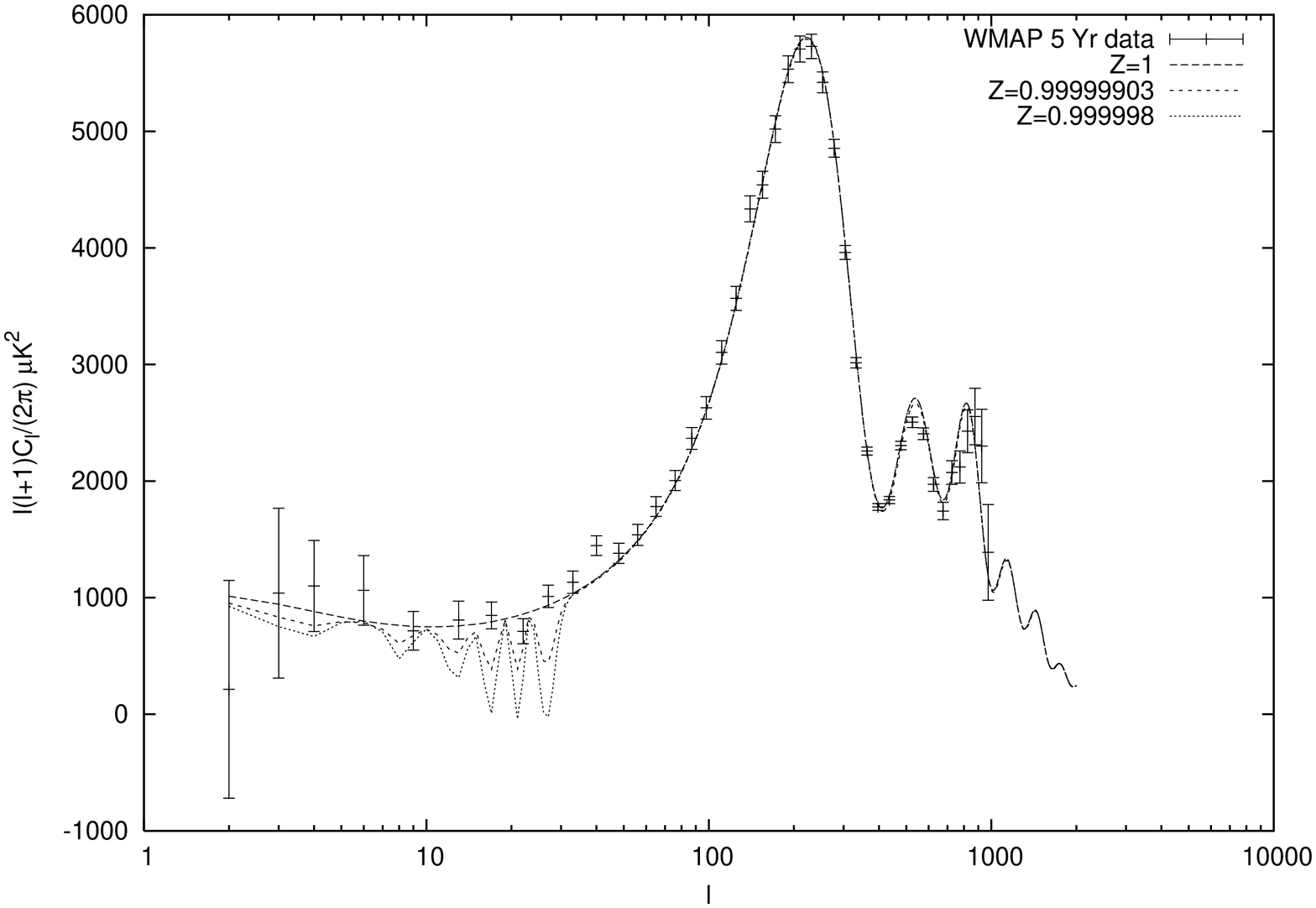}
\caption{The TT angular spectrum for the inflaton as a
composite particle.}
\label{fig_composite}
}
In Fig~(\ref{fig_composite}) we plot the TT angular spectrum for the
case of a composite inflaton. The parameters used for these plots
are $H=10^{13}$ GeV, $m_{\varphi}=1.0\times 10^{12}$ GeV,
$m_0=3.0\times 10^{12}$ GeV, $s_0=1.0\times 10^{11}$ GeV,
$\Lambda=1.0\times 10^{13}$ GeV, $N=3$ and for $1-Z=9.7 \times
10^{-7}$ and $1-Z=2.0\times 10^{-6}$ we take $f_{\varphi}=1.4\times
10^{16}$ GeV, $\epsilon=3.92\times 10^{-6}$ and $f_{\varphi}=1.0\times
10^{16}$ GeV, $\epsilon=3.87\times10^{-6}$ respectively. We find
that there are oscillatory features in the power spectrum at $l=30$.

Analysis of WMAP data by several groups \cite{oscillations} suggests
that the power spectrum may have such oscillatory features.  We have
given the plot for some plausible values of the parameters.  A
detailed fit of the parameters with WMAP data using COSMO-MC
\cite{Cosmo} will be followed up in a forth-coming paper.


\section{Conclusion}

We have derived a general formula for incorporating short-range
interactions in the two-point correlation functions and the power
spectrum by use of the K\"{a}ll\'{e}n-Lehmann spectral function of
flat spacetime. This method is useful if the short wavelength limit of
the mode functions are plane wave states
$\frac{1}{\sqrt{2k}}e^{-ik\eta}$, follows in the quasi de-Sitter
inflation by the assumption of the Bunch-Davis boundary conditions and
there exists a complete orthonormal set of mode functions of the free
theory in curved spacetime which is true in the quasi de-Sitter space
relevant for inflation power spectrum calculation. In interacting
inflaton models like the ones studied in this paper we find that there
are more interesting variations in the power spectrum due to the
modification of the propagators which do not affect the slow roll
parameters. We apply our formulation to study inflation with decaying
and composite inflatons. We find that the decay of the inflaton
results in the suppression of long distance correlations and thereby a
loss of the quadrupole anisotropy \cite{decay-ref}.  This may be
related to the observation of low quadrupole power by WMAP
\cite{wmap}.

When the inflaton is taken as a composite of two fermions the power
spectrum displays even more interesting features like oscillations. An
examination of the WMAP data by wavelet analysis and by the cosmic
inversion method reveals that the data may have such features
\cite{oscillations}.
\acknowledgments

We have used the public domain code CAMB \cite{camb} to generate the
$C_l$ plots in Fig.~(\ref{fig_decay}) and Fig.~(\ref{fig_composite}).
We like to thank the anonymous referee for his useful suggestions.
\appendix
\section{Propagator for interacting scalar field in Minkowski space using KL representation}
\label{KL}
KL representation is a non-perturbative way to derive propagator for
interacting fields. Here a brief description of deriving the Feynman
propagator and the Wightman function of interacting real scalar fields
is being discussed. A more detailed derivation can be found in
\cite{weinberg}. Considering a generic scalar field $\Phi(x)$, the
vacuum expectation value of the time-ordered product $\langle 0|
\mathcal{T}\lbrace\Phi(x)\Phi(y)\rbrace|0\rangle$ gives the complete
Feynman propagator for the scalar field in Fourier space
\begin{eqnarray}
-i\Delta^{\prime}(p)=\int d^4x\exp[ip\cdot(x-y)]\langle 0|
\mathcal{T}\lbrace\Phi(x)\Phi(y)\rbrace|0\rangle,
\label{propagator}
\end{eqnarray}
while the vacuum expectation value of product of two scalar fields is
known as the Wightman function 
\begin{eqnarray}
W^{\prime}(x-y)=\langle0|\Phi(x)\Phi(y)|0\rangle.
\label{wightman}
\end{eqnarray}
Inserting a complete set of momentum eigenstates in between
the two field operators the vacuum expectation value of
$\Phi(x)\Phi(y)$ can be written as
\begin{eqnarray}
\langle0|\Phi(x)\Phi(y)|0\rangle=\sum_n\langle0|\Phi(x)|n\rangle\langle n|\Phi(y)|0\rangle.
\label{phi-phi}
\end{eqnarray}
Translational invariance in Minkowski space yields
\begin{eqnarray}
\Phi(x)=\exp(ip\cdot x)\Phi(0)\exp(-ip\cdot x),
\end{eqnarray}
where
\begin{eqnarray}
\langle0|\Phi(x)|n\rangle&=&\exp(-ip_n\cdot x)\langle0|\Phi(0)|n\rangle
\nonumber \\
\langle n|\Phi(y)|0\rangle&=&\exp(ip_n\cdot y)\langle n|\Phi(0)|0\rangle.
\label{mode}
\end{eqnarray}
In Minkowski space therefore Eq.~(\ref{phi-phi}) can be written as
\begin{eqnarray}
\langle0|\Phi(x)\Phi(y)|0\rangle=\sum_n\exp(-ip_n\cdot(x-y))|\langle0|\Phi(0)|n\rangle|^2.
\end{eqnarray}
$|\langle0|\Phi(0)|n\rangle|^2$ encapsulating the interacting features
of the scalar field can be replaced by a spectral function
$\rho(q^2)$ defined as
\begin{eqnarray}
\theta(q^0)\rho(q^2)=(2\pi)^3\sum_n\delta^4(q-p_n)|\langle0|\Phi(0)|n\rangle|^2.
\label{sf}
\end{eqnarray}
The spectral function $\rho(q^2)$ is a function of $q^2$ due to
Lorentz invariance and is real, positive and vanishes for
$q^2<0$. With this definition of spectral function Eq.~(\ref{phi-phi})
can be expressed as
\begin{eqnarray}
\langle0|\Phi(x)\Phi(y)|0\rangle=\int_0^{\infty}d\sigma^2\rho(\sigma^2)\Delta(x-y;\sigma^2),
\label{phi-phi-1}
\end{eqnarray}
where
\begin{eqnarray}
\Delta(x-y;\sigma^2)=\frac{1}{\left(2\pi\right)^3}\int d^4q\exp[-iq\cdot(x-y)]\theta(q^0)\delta(q^2-\sigma^2),
\label{del}
\end{eqnarray}
and $\sigma$ is known as the mass parameter. Similarly one can find
\begin{eqnarray}
\langle0|\Phi(y)\Phi(x)|0\rangle=\int_0^{\infty}d\sigma^2\rho(\sigma^2)\Delta(y-x;\sigma^2),
\label{phi-phi-2}
\end{eqnarray}
where
\begin{eqnarray}
\Delta(y-x;\sigma^2)=\frac{1}{\left(2\pi\right)^3}\int d^4q\exp[-iq\cdot(y-x)]\theta(q^0)\delta(q^2+\sigma^2).
\end{eqnarray}

\subsection{Feynman propagator for interacting scalar field}
\label{KL-f}
The vacuum expectation value of two time-ordered field
operators is
\begin{eqnarray}
\langle0|\mathcal{T}\lbrace \Phi(x)\Phi(y)\rbrace|0\rangle=\Theta(x_0-y_0)\langle0|\Phi(x)\Phi(y)|0\rangle+\Theta(y_0-x_0)\langle0|\Phi(y)\Phi(x)|0\rangle.
\label{time-order}
\end{eqnarray}
Inserting Eq.~(\ref{time-order}), Eq.~(\ref{phi-phi-1}) and
Eq.~(\ref{phi-phi-2}) in Eq.~(\ref{propagator}) gives the propagator
for interacting scalar field as
\begin{eqnarray}
-i\Delta^{\left({\rm int}\right)}(p)=-i\int d^4x\exp[ip\cdot(x-y)]\int_0^{\infty}d\sigma^2\rho(\sigma^2)\Delta_F(x-y;\sigma^2),
\end{eqnarray}
where the Feynman propagator $\Delta_F(x-y;\sigma^2)$ for the scalar field is
\begin{eqnarray}
-i\Delta_F(x-y;\sigma^2)&=&\Theta(x_0-y_0)\Delta(x-y;\sigma^2)+\Theta(y_0-x_0)\Delta(y-x;\sigma^2)\nonumber \\
&=&\frac{-i}{\left(2\pi\right)^4}\int d^4q\exp[-iq\cdot(x-y)]\frac{1}{q^2-\sigma^2-i\varepsilon}.
\end{eqnarray}
To derive the last equality the form of the step function
\begin{eqnarray}
\Theta(t)=-\frac{1}{2\pi i}\lim_{\varepsilon\rightarrow 0}\int_{-\infty}^{+\infty}\frac{e^{-ist}}{s+i\varepsilon}ds
\end{eqnarray}
is to be used. This yields the form of the full propagator for the
interacting scalar field in terms of the spectral function as
\begin{eqnarray}
\Delta^{\left({\rm int}\right)}(p)=\int_0^{\infty}d\sigma^2\rho(\sigma^2)\frac{1}{p^2-\sigma^2+i\varepsilon}.
\end{eqnarray}
$\frac{1}{p^2-\sigma^2+i\varepsilon}$ can be recognized as the propagator
for a free scalar field with the mass $m$ of the scalar field replaced
by the mass parameter $\sigma$. Hence one can write the above equation
as
\begin{eqnarray}
\Delta^{\left({\rm int}\right)}(p)=\int_0^{\infty}d\sigma^2\rho(\sigma^2)\Delta^0(p;\sigma^2),
\end{eqnarray}
where $\Delta^0(p;\sigma^2)\equiv\frac{1}{p^2-\sigma^2+i\varepsilon}$ is the
free propagator of the scalar field.
\subsection{Wightman function for interacting scalar field}
\label{KL-w}
For a free scalar field with mass $m$ the Wightman function defined in
Eq.~(\ref{wightman}) is
\begin{eqnarray}
W^0(x-y)=\frac{1}{\left(2\pi\right)^3}\int
\frac{d^3k}{2\omega_k}e^{-i\omega_k\left(x_0-y_0\right)+i{\mathbf
    k}\cdot({\mathbf x}-{\mathbf y})}.
\label{w0}
\end{eqnarray}
where $\omega_k\equiv\sqrt{{\mathbf k}^2+m^2}$. For the interacting scalar
field the Wightman function can be derived using Eq.~(\ref{phi-phi-1})
which turns out to be
\begin{eqnarray}
W^{\left({\rm int}\right)}(x-y)=\int_0^{\infty}d\sigma^2\rho(\sigma^2)\Delta(x-y;\sigma^2),
\label{wf}
\end{eqnarray}
where $\Delta(x-y;\sigma^2)$ given in Eq.~(\ref{del}) can be written as
\begin{eqnarray}
\Delta(x-y;\sigma^2)=\frac{1}{\left(2\pi\right)^3}\int
\frac{d^3q}{2\omega_q}e^{-i\omega_q\left(x_0-y_0\right)+i{\mathbf
    q}\cdot({\mathbf x}-{\mathbf y})}.
\end{eqnarray}
This can be identified as the Wightman function for the free scalar
field given in Eq.~(\ref{w0}) where the mass $m$ of the scalar field
is replaced by the mass parameter $\sigma$ and
$\omega_q\equiv\sqrt{{\mathbf q}^2+\sigma^2}$ and hence Eq.~(\ref{wf})
can be written as
\begin{eqnarray}
W^{\left({\rm int}\right)}(x-y)=\int_0^{\infty}d\sigma^2\rho(\sigma^2)W^0(x-y;\sigma^2).
\end{eqnarray}
The equal-time Wightman function $(x_0=y_0)$ for the
interacting scalar field has the form
\begin{eqnarray}
W^{\left({\rm int}\right)}_{\rm
  ET}(x-y)&=&\frac{1}{\left(2\pi\right)^3}\int_0^{\infty}d\sigma^2\rho(\sigma^2)\int
\frac{d^3q}{2\omega_q}e^{i{\mathbf q}\cdot({\mathbf x}-{\mathbf y})}\nonumber\\
&=&\int_0^{\infty}d\sigma^2\rho(\sigma^2)W^0_{\rm ET}(x-y;\sigma^2).
\end{eqnarray}
\section{Power spectrum of free  scalar field}
\label{power}

The useful quantity to characterize the properties of quantum
fluctuations in the inflaton field is the power spectrum which is the
variance (two-point correlation function) of these fluctuations. The
quantum fluctuations of inflaton field cab be expanded in Fourier
modes as
\begin{equation}
\delta\phi({\mathbf x},t)=\int \frac{d^3k}{\left(2\pi\right)^{\frac32}}e^{i{\mathbf k}\cdot{\mathbf x}}\delta\phi_{\mathbf k}(t),
\end{equation}
which satisfy the Klein-Gordon equation in momentum space
\begin{eqnarray}
\delta\ddot{\phi}_{\mathbf k}+3H\delta\dot{\phi}_{\mathbf k}+\left(\frac{k^2}{a^2}+m_\phi^2\right)\delta\phi_{\mathbf k}=0,
\label{KG}
\end{eqnarray}
where dot represents derivative with respect to cosmic time $t$.  The
power spectrum for these fluctuations is defined as
\begin{eqnarray}
\delta^3({\mathbf k_1}-{\mathbf k_2}){\mathcal P}^{\left(0\right)}_{\delta\phi}(k)\equiv \frac{k^3}{2\pi^2}\langle0|\delta\phi_{\mathbf k_1}\delta\phi_{\mathbf k_2}|0\rangle,
\label{pspectrum}
\end{eqnarray}
and hence the variance of the perturbations at equal time in term of
power spectrum turns out to be
\begin{eqnarray}
\langle0|(\delta\phi({\mathbf x},t))^2|0\rangle=\int\frac{dk}{k}{\mathcal P}^{\left(0\right)}_{\delta\phi}(k).
\label{var}
\end{eqnarray}
Defining $\delta\phi_{\mathbf k}=\frac{\delta\chi_{\mathbf
    k}}{a(\eta)},$ the equation of motion satisfied by
$\delta\chi_{\mathbf k}$ can be derived from Eq.~(\ref{KG}) as
\begin{eqnarray}
\delta\chi_{\mathbf k}^{\prime\prime}+\left(k^2-\frac{1}{\eta^2}\left(\nu_\phi^2-\frac14\right)\right)\delta\chi_{\mathbf k}&=&0,
\end{eqnarray}
where the prime denotes derivative with respect to conformal time
$\eta$ and
$\nu_\phi^2\equiv\left(\frac94-\frac{m_\phi^2}{H^2}\right)$. This
equation for real $\nu_\phi$ (i.e. $\frac{m_\phi}{H}<\frac32$) has the
following solution
\begin{eqnarray}
\delta\chi_{\mathbf k}=\sqrt{-\eta}\left[c_1(k){\rm H}^{\left(1\right)}_{\nu_\phi}(-k\eta)+c_2(k){\rm H}^{\left(2\right)}_{\nu_\phi}(-k\eta)\right],
\end{eqnarray}
where ${\rm H}^{\left(1\right)}_{\nu_\phi}$ and ${\rm
  H}^{\left(2\right)}_{\nu_\phi}$ are the Hankel functions of the
first and second kind respectively having the form
\begin{eqnarray}
{\rm H}^{\left(1\right)}_{\nu_\phi}(x\gg1)&\sim& \sqrt{\frac{2}{\pi x}}e^{i\left(x-\frac{\pi}{2}\nu_\phi-\frac{\pi}{4}\right)}\nonumber \\
{\rm H}^{\left(2\right)}_{\nu_\phi}(x\gg1)&\sim& \sqrt{\frac{2}{\pi x}}e^{-i\left(x-\frac{\pi}{2}\nu_\phi-\frac{\pi}{4}\right)}.
\end{eqnarray}
Imposing that the modes of these scalar fluctuations inside the Hubble
radius $(k\gg aH)$ have the plane wave solution
$\frac{e^{-ik\eta}}{\sqrt{2k}}$ yields
$c_1(k)=\frac{\sqrt{\pi}}{2}e^{i\left(\nu_\phi+\frac12\right)\frac{\pi}{2}}$
and $c_2(k)=0$. On superhorizon scale $(k\ll aH)$ using the asymptotic
behaviour of the Hankel function
\begin{eqnarray}
{\rm
  H}^{\left(1\right)}_{\nu_\phi}(x\ll1)\sim\sqrt{\frac{2}{\pi}}e^{-i\frac\pi
  2}2^{\left(\nu_\phi-\frac32\right)}\frac{\Gamma\left(\nu_\phi\right)}{\Gamma\left(\frac32\right)}x^{-\nu_\phi},
\end{eqnarray}
the solution for $\delta\phi_{\mathbf k}$ turns out to be
\begin{eqnarray}
\left|\delta\phi_{\mathbf k}\right|\simeq\frac{H}{\sqrt{2k^3}}\left(\frac{k}{2aH}\right)^{\frac32-\nu_\phi},
\label{sol-phi}
\end{eqnarray}
and hence the power spectrum defined in Eq.~(\ref{pspectrum}) for this
light scalar field $(m_\phi<H)$ will be
\begin{eqnarray}
{\mathcal
P}^{\left(0\right)}_{\delta\phi}(k)=\frac{H^2}{4\pi^2}\left(\frac{k}{2aH}\right)^{\frac23\frac{m_\phi^2}{H^2}}.
\label{light}
\end{eqnarray}
For very massive scalar field $(m_\phi>H)$ i.e. for imaginary
$\nu_\phi$ the form of power spectrum can be derived as

\begin{eqnarray}
{\mathcal
P}^{\left(0\right)}_{\delta\phi}(k)\simeq\frac{H^2}{4\pi^2}\left(\frac{H}{m_\phi}\right)\left(\frac{k}{aH}\right)^3,
\label{heavy}
\end{eqnarray}
which is suppressed by the ratio $\left(\frac{H}{m_\phi}\right)$ and
hence highly damped at large wavelengths.

These quantum fluctuations in inflaton field generates fluctuation in
the metric which is coupled to it through Einstein's equation. The
perturbed FRW metric has the form (considering only scalar
perturbations)
\begin{eqnarray}
\tilde{g}_{\mu\nu}=a^2(\eta)\left(
\begin{array}{ccc}
1+2A & & 0 \\
0 & & -(1-2\psi)\delta_{ij}
\end{array}
\right),
\end{eqnarray}
where the quantity $\psi$ is known as the curvature perturbation. The
gauge invariant quantity formed out of this perturbation is known as
the comoving curvature perturbation and defined as
\begin{eqnarray}
{\mathcal R}=\psi+H\frac{\delta\phi}{\dot{\phi}}.
\end{eqnarray}
The CMB anisotropy spectrum is determined by the power spectrum of
this comoving curvature perturbation which is related to the power
spectrum of scalar perturbation as
\begin{eqnarray}
{\cal P}_{\cal R}(k)=\frac{1}{2m_{\rm Pl}^2\epsilon}{\mathcal P}^{\left(0\right)}_{\delta\phi}(k),
\label{curvature}
\end{eqnarray}
where $\epsilon$ is the slow-roll parameter and $m_{\rm
  Pl}=\frac{1}{\sqrt{8\pi G}}$, $G$ being the Newton's Gravitational
constant.


\end{document}